\documentclass[aps,reprint,prl]{revtex4-2}

\usepackage{graphicx,comment}
\usepackage{dcolumn}
\usepackage{bm}
\usepackage{amsmath,amssymb,amsfonts}
\usepackage[english]{babel}
\usepackage{natbib}
\usepackage{braket}
\usepackage{float}
\usepackage[caption=false]{subfig}
\usepackage[dvipsnames]{xcolor}
\usepackage{hyperref}

\begin{document}
\title{{Tuning Quantum States at Chirality-Reversed Planar Interface in  Weyl Semimetals using an Interstitial Layer}}

\date{\today}

\author{Eklavya Thareja}
\email{ethareja@usf.edu}
\affiliation{Department of Physics, University of South Florida, Tampa, FL 33620}

\author{Gina Pantano}
\affiliation{Department of Physics, University of South Florida, Tampa, FL 33620}

\author{Ilya Vekhter}
\affiliation{Department of Physics and Astronomy, Louisiana State University, Baton Rouge, LA 70803}

\author{Jacob Gayles}
\affiliation{Department of Physics, University of South Florida, Tampa, FL 33620}

\begin{abstract}
The electronic band structure of Weyl semimetals possesses pairs of linear band crossings, called Weyl nodes, characterized by opposite chirality charges associated with each node. The momentum space position of the nodes can reverse across a planar interface and these host Fermi-arc-like bound states, in addition to scattering states. We show that a magnetic interstitial layer can tune these states in three distinct ways. The electrostatic potential and one of the in-plane magnetic potential components control the shape of the bound state Fermi-arcs. For moderate values of the same in-plane magnetic potential electrons are spin-filtered across the interface, while both the in-plane magnetic components and the electrostatic potential control the transmission of electrons. The ratio of in-plane to out-of-plane magnetic components can be used to turn on or turn off the magnetic potential effects, since the latter does not affect the interface states. The tunability arises from spin-momentum locking and chirality reversal at the interface. Thus, the effects can mix or interchange depending on the specific material but the states will remain tunable.
\end{abstract}

\maketitle


\paragraph{Introduction.} Materials with chiral electronic states have been at the forefront of applied physics proposals such as dissipation-less electron transport \cite{Hu_2019}, racetrack memory \cite{Parkin_2008}, domain wall (DW) logic \cite{Allwood_2005}, and chiral molecule sensing \cite{Cavicchi_2024,Moreno_2024}. These materials can host massless Dirac quasiparticles and have generated substantial interest over the past two decades as their transport properties are generally robust against disorder \cite{Review_hasan_2010, Review_Qi_2011, Review_Wehling_2014, Weyl_Vishwanath_review, Manna_2018, Review_Lv_2021}. This is due to topological features of the electronic wavefunction that dominate transport when they are close to the Fermi surface of these materials. 
Weyl semimetals (WSMs) have linear crossings in the bulk dispersion, called Weyl nodes, that appear in pairs. These nodes in the WSM act as sources and sinks of Berry curvature, a momentum space analog of the magnetic field. The Berry curvature results in a chiral charge ($\chi$) associated with each node that is quantized to values $\chi = \pm1$ in the simplest cases. The chiral node pair in momentum space can give rise to surface Fermi-arcs \cite{Xu_2015,Morali_2019,Vergniory_2019,Sakhya_2023,li_2023,Robredo_2024}. 
WSMs are known to exhibit electromagnetic effects such as chiral magnetic effect \cite{Li_2016} and negative magnetoresistance \cite{Huang_2015,Zhang_2016}, that result from these features.

\begin{figure}[t]
\includegraphics[width=0.8\columnwidth]{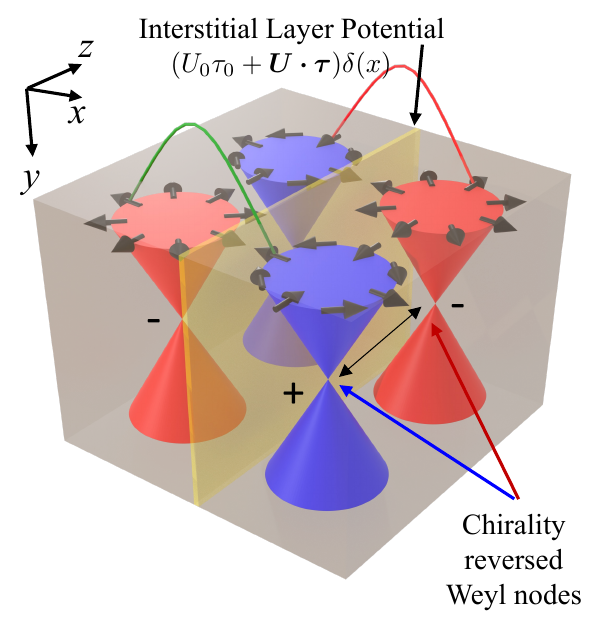}
\caption{Chirality-reversed planar interface (CRPI) with interstitial layer in a magnetic Weyl semimetal with spin texture of the electronic states indicated by the arrows on the cones. 
}
\label{fig:model}
\end{figure}


Planar defects such as twin boundaries, which were recently observed in CoSi \cite{Mathur_2023}, can result in the exchange of chirality between the nodes. In magnetic WSMs, a reversal in magnetization across a defect can also reverse node chirality, as shown in Fig. \ref{fig:model}. A similar situation can arise in polycrystalline materials or the presence of microstructures. The electronic states on these defects are analogous to DW states \cite{Araki_2016, Araki_2018}, that promise robustness against disorder due helicity protection \cite{Kobayashi_2018}. Other resultant effects such as quantum oscillations \cite{Kaushik_2022,Chaou_2024} and chiral magnetic effect \cite{Chaou_2023} from interface Fermi-arcs have also been proposed. Thus, WSMs can host physically interesting bound states at the crystallographic or magnetic domain boundaries. 

In this letter, we show that an interstitial layer at the chirality-reversed planar interface in WSMs  allows for a high degree of control over the electronic states. Their tunability is a direct consequence of spin-momentum locking (SML) differences between nodes of different chirality. We show that Fermi-arc-like states bound to the interface exist regardless of the magnetic or nonmagnetic interstitial layer potentials. The electrostatic potential and the in-plane magnetic component perpendicular to the Weyl node split direction determine the shape of the Fermi arc. The in-plane magnetic component parallel to the Weyl node split direction determines the asymmetry local density of states across the interface. On the other hand, the out-of-plane magnetic component does not affect the bound states due to presence of SML.

Scattering properties also depend on the electrostatic and in-plane magnetic potentials but not on the out-of-plane component. The in-plane magnetic components control the opaqueness of the interface, and the in-plane component perpendicular to node split direction filters spin for moderate values, compared to hopping potential. Hence, our work shows high tunability of electronic states near a chirality-reversed planar interface (CRPI) using an interstitial layer. 

Since SML can vary based on the intrinsic material properties, the effects of different potential components can mix or interchange but remain tunable. Our work can universally be extended to the chirality-reversal in inversion breaking (non-magnetic) WSMs and similar tunable states will arise at these planar interfaces.

\paragraph{Model.} Weyl nodes in the bulk band structure act as sources of Berry curvature and are key topological features in WSMs. The vanishing bulk density of states at the nodes makes excitation behavior near the nodes critical for electronic transport. The smallest number of nodes is preferable to highlight the novel physics; therefore, we consider a minimal model for time-reversal-breaking WSMs with two nodes, opposite in chirality. Our simplification is generic and applicable to systems such as K$_2$Mn$_3$(AsO$_4$)$_3$ containing precisely two Weyl nodes \cite{Nie_2022} and other compounds \cite{Kushwaha_2018} with a small number of nodes near the Fermi energy such as Co$_3$Sn$_2$S$_2$ \cite{Liu_2019}. 
We utilize a previously proposed \cite{Trivedi_2017,Weyl_Vishwanath_review} simple cubic tight-binding model with nodes at $k_z = \pm k_0$: $H_{TB}({\bm k}) = t\tau_z(2+\cos k_0a -\cos k_xa -\cos k_y a -\cos k_z a) + t \tau_x\sin k_xa  + t \tau_y\sin k_ya$,
where $t$, $a$ and $\bm{\tau}$ are the hopping parameter, lattice constant and Pauli matrices in the spin subspace. The magnitude of magnetization determines the chiral splitting between the nodes, $2k_0$, and direction is representative of easy-axis anisotropy in real materials. 

Expanding around the nodes using $\bm{q} = \bm{k} \mp k_0 \hat{k_z}$, captures the low-energy behavior. Then, $H_0({\bm q}) = ta(-n \tau_zq_z \sin k_0a + \tau_y q_y + \tau_x q_x) = \bm{v_n\cdot\tau}$,
where $n=\pm1$ correspond to nodes at $k_z =\pm k_0$. We show $y-z$ projected spin texture of the nodes in Fig. \ref{fig:model} for $x<0$. The chirality of the Weyl nodes, $\chi = \text{sgn}(v_x\cdot (v_y \times v_z))$, can be switched by flipping the SML along either one or all three directions. This model has SML flipped along $z$-axis.


{We model a uniform and sharp potential from the interstitial layer as $U = (U_0 \tau_0 + \bm{U\cdot \tau})\delta (x)$, which is valid when potential width is $\ll 1/(2k_0)$, and thus inter-nodal scattering can be ignored.} The Hamiltonian,
\begin{multline}
\label{eq:H_D}
H_{CRPI}(\bm{q},x) = \text{sgn}(x) t a(n \tau_zq_z\sin k_0a) \\+ (U_0 \tau_0 + \bm{U\cdot \tau})\delta (x) +\tau_y q_y + \tau_x \hat{q}_x,
\end{multline}
describes WSM with CRPI located at $x=0$. Here $\hat{q}_x = -i\partial/\partial x$ due to broken translation symmetry along $x$. 
We can assume $t>0$ and $a=1$ without loss of generality.

\paragraph{Boundary Conditions.} {The wavefunction is discontinuous at the interstitial layer for a Dirac equation with a $\delta$-function potential because it is linear in $\hat{q}_x$.} 
We directly integrate the hamiltonian in Eq. \eqref{eq:H_D}, to find boundary condition at $x=0$ \cite{Levitov_2018,Thareja_2023},\begin{align}
\label{eq:BC}
\psi(0+) = e^{-\sum_{j=0,x,y,z} iU_j\tau_x \tau_j/t} \psi(0-) = M \psi(0-).
\end{align}
The out-of-plane magnetic potential, $U_x\tau_x$, results in $M_x = e^{-i U_x \tau_0/t}$, which is a constant phase along the interface. Thus, observables are unchanged across the interface. The electrostatic potential $U_0$ appears as a rotation in the interface plane i.e. $M_0 = e^{ -iU_0 \tau_x/t}$. Interestingly, in-plane potential, $U_y\tau_y$, results in a boundary condition $M_y = e^{-U_y \tau_z/t}$, that can be rewritten as $M_y = e^{U_y/t} P_z(+) + e^{-U_y/t}P_z(-)$, where $P_i(\pm)$ is projection operator along $i$-axis. Thus, for a large and positive (negative) $U_y$ the spinor is projected along (against) $z$-axis. Similarly, the presence of the other in-plane potential, $U_z\tau_z$, results in $M_z = e^{-U_z \tau_y/t} = e^{U_z/t} P_y(-) + e^{-U_z/t} P_y(+) $, where spinor projections are along $y$-axis. Change in the form of SML can mix or switch the roles played by each potential, but these distinct effects that produce the tunability of states will still be present and encoded in corresponding $M$-matrices.

\begin{figure}
\includegraphics[width=\columnwidth]{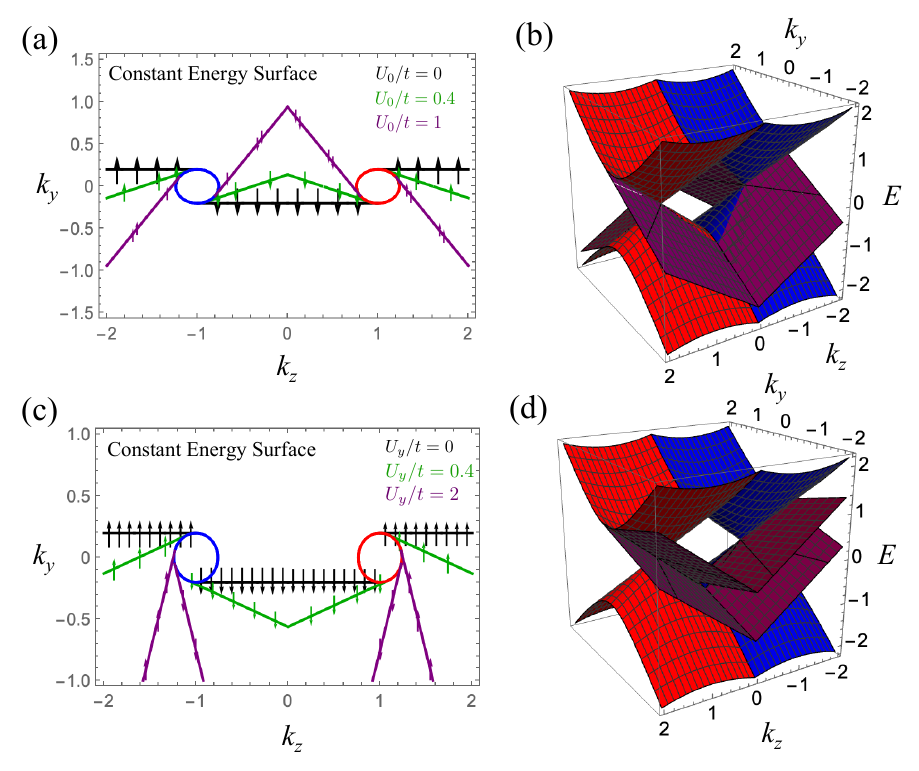}
\caption{(a),(c) Constant energy cross section at $E = 0.2 t$ for different potentials. (b),(d) Weyl nodes and excited states near the nodes (blue and red) and bound state dispersion (purple) along the CRPI for (b) $U_0 = t$  and (d) $U_y = 2t$.}
\label{fig:bound_states}
\end{figure}

\paragraph{Bound States.} WSMs possess surface Fermi arcs that connect Weyl nodes. DWs between materials with complete reversal of spin-momentum texture $\bm{v_k \cdot\tau } \rightarrow -\bm{v_k \cdot\tau}$ also host similar states \cite{Araki_2016,Araki_2018}. We show that bound states are also present when chirality is reversed by changing only $v_z \rightarrow -v_z$, thus their existence is more general.
The spatial dependence of the wavefunction is $\psi_b \propto e^{iq_y y} e^{i q_z z} e^{\mp\lambda x}$, see SM for full form \cite{SM}. The energy and inverse localization length are $E_b = \epsilon t \sqrt{q_y^2 + q_z^2 \sin^2 k_0} \sin\alpha$ and $\lambda = \sqrt{q_y^2 + q_z^2 \sin^2 k_0} \cos\alpha$, respectively, where $\epsilon = +1$ (-1) indicates energy is above (below) Weyl nodes, and along with $0<\alpha<\pi$, is determined by imposing Eq. \eqref{eq:BC}.

Without the interstitial layer potential, which corresponds to having a sharp collinear DW \cite{Araki_2016}, $\sin\alpha =  (-1)^m \cos\beta$ when $\lambda>0$ i.e. $\cos\alpha =  (-1)^m \sin\beta>0$. For the low-energy model, we solve separately near each node. The Fermi arcs are then formed by combining the solution for node $n=1$ for all $k_z>0$ and node $n=-1$ for all $k_z<0$. Two branches of solutions exist, one for odd $m$ and other for even $m$. The energy upon substitution of $\beta$ is $E = \epsilon(-1)^m k_y$ where $m$ is chosen such that $(-1)^m q_z\sin k_0 > 0$, so the solutions are dispersionless along $k_z$. The constant energy surface, in Fig. \ref{fig:bound_states}(a), shows one of the branches of the interface bound states connects fermi surfaces near Weyl nodes, similar to surface Fermi-arcs. In contrast, the other branch meets across the Brillouin zone which is pushed to infinity in a continuum model. The localization length is $1/\lambda = 1/|q_z \sin k_0|$, thus, states further away from the nodes are more localized than the states closer to the nodes. 

In the presence of electrostatic potential at the interface, the solutions are $\sin\alpha =  (-1)^m \cos(\beta-nU_0/t)$ and $\cos\alpha =  (-1)^m \sin(\beta-nU_0/t)$, 
where $m$ has to be chosen to satisfy $\lambda = \sqrt{q_y^2 + q_z^2 \sin^2 k_0} \cos\alpha >0$. The effect of the electrostatic potential is a rotation of the arcs in the plane of the interface, see Fig. \ref{fig:bound_states}(a). The states now disperse along both $k_y$ and $k_z$, see Fig. \ref{fig:bound_states}(b). The oscillatory behavior of the dispersion with $U_0$ is analogous to oscillations from Fabry-Perot-like interference that occurs for finite-width interlayer potential in 2D Dirac materials \cite{Katsnelson2006,Shytov_2008,Thareja_2020}. Without the potential, the momentum space spin texture of the states points along $y$. The rotation of the spinor part of the wavefunction generates oscillations in $\braket{\tau_y}$, and other physical properties, with period $2\pi t/U_0$.


The effect of magnetic potentials from the interstitial layer is as follows. First, the potential $U_x\tau_x$ does not affect the interface states. This agrees with the argument that a Bloch DW in WSMs can be gauge transformed into a collinear DW, which corresponds to no interstitial layer case \cite{Araki_2016}. Second, for $U_y \tau_y$ potential, 
the arcs attract each other, as shown in Fig. \ref{fig:bound_states}(c) and \ref{fig:bound_states}(d). As $U_y/t \rightarrow \infty$, the two arcs merge. Third, the magnetic potential $U_z\tau_z$ does not affect the shape of fermi arcs but changes their spatial distribution, as described in the LDOS section below. This establishes that only the in-plane component perpendicular to the direction of chiral splitting, $U_y\tau_y$, and electrostatic potential, $U_0\tau_0$, determine the shape of the fermi arcs near the Weyl nodes. 

\begin{figure}[H]
\includegraphics[width=\columnwidth]{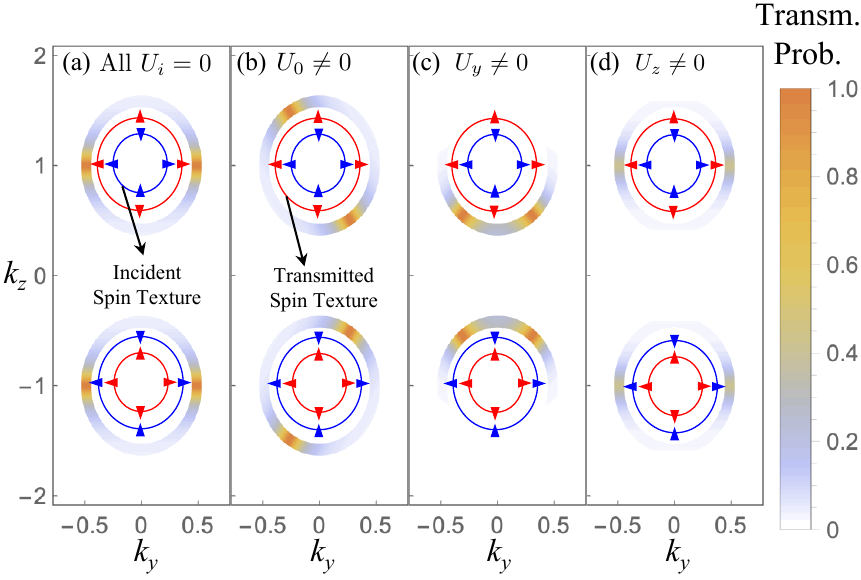}
\caption{Transmission probability of electrons through the CRPI with variation in the in-plane momenta (a) without or (b-d) with interstitial layer potentials. Inner (outer) circle arrows represent spin texture of the incident (transmitted) electron. The incident momentum component perpendicular to the interface $k_x = 0.1$, energy $E = 0.5 t$, and interface potentials (b) $U_0 = t$, (c) $U_y = t$ and (d) $U_z = t$.}
\label{fig:transm_prob}
\end{figure}

\begin{widetext}

\begin{figure}
\includegraphics[width=0.9\columnwidth]{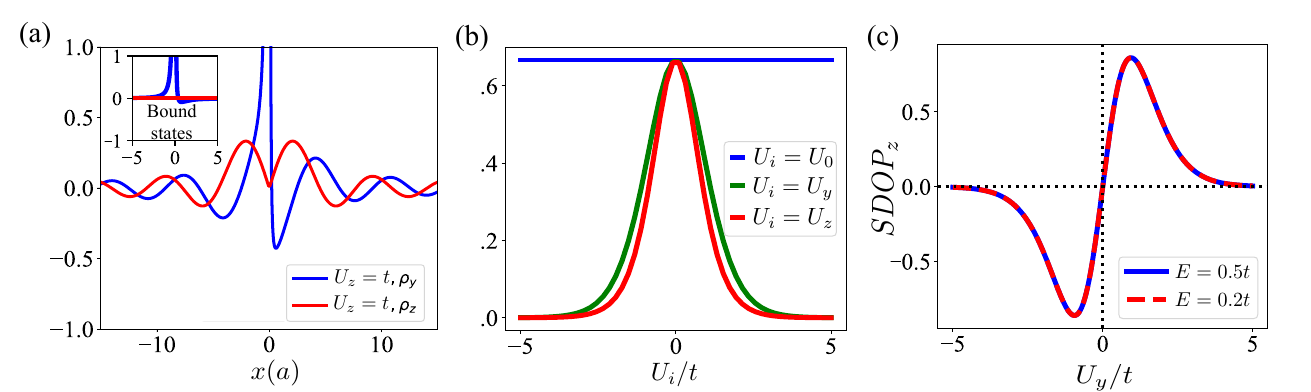}
\caption{(a) Spin-projected LDOS in presence of magnetic potential $U_z\tau_z$ at $E = 0.5 t$. Inset: Asymmetric spin-resolved LDOS across the interface. (b) Longitudinal conductance across CRPI as function of electrostatic and in-plane magnetic potentials with $E = 0.5 t$. (c) Variation of Spin Degree of Polarization ($SDOP_z$) with magnetic potential $U_y$. }
\label{fig:ldos}
\end{figure}
\end{widetext}

\paragraph{Scattering States.} There are two independent processes, assuming no internode scattering. The process 1 (2) refers to electrons incident from $x\rightarrow -\infty$ ($x\rightarrow \infty$). The wavefunction for process 1 is of the form $\psi_1^{(s)}(x<0) = (\psi_1^{(i)} e^{iq_x x} + r_1 \psi_1^{(r)} e^{-iq_x x})e^{iq_y y}e^{i (q_z +nk_0) z}$ and $\psi_1^{(s)}(x>0) = t_1 \psi_1^{(t)} e^{iq_x x}e^{iq_y y}e^{i (q_z +nk_0) z}$. The full wavefunctions are in the supplementary material (SM) \cite{SM}.
The maxima of transmission probability, $|t_1|^2 = 1-|r_1|^2$, is determined by two factors: (i) the spinor structure matching between the incident and the transmitted electron, and (ii) the effect of interstitial layer potential captured by $M$-matrix. 


The spin structure of an incident and a transmitted electron is shown in Fig. \ref{fig:model}. In absence of interlayer potential, $M$ reduces to $\tau_0$ and spinor structure is matched only for $q_z = 0$ i.e. $k_z = \pm k_0$ and $k_y = q_y = \pm \sqrt{E^2-q_x^2}$. In the presence of electrostatic potential, $M$-matrix rotates the spinor in $y-z$ plane, leading to a corresponding rotation of the transmission probability by an angle $U_0/t$, see Figs. \ref{fig:transm_prob}(a) and \ref{fig:transm_prob}(b). Since the spinor texture between the nodes at $k_z = +k_0$ and $k_z = -k_0$ is switched between the two sides, the resultant transmission maxima rotate with opposite handedness for the two nodes.


The magnetic potential $U_x \tau_x$ does not cause any additional scattering other than one caused by chirality reversal across the interface. For $U_y\tau_y$ potential the boundary matrix satisfies $M^\dagger_y \tau_{x,y} M_y = \tau_{x,y}$, thus the potential leaves $\braket{\tau_{x,y}}$ unchanged. The chirality reversal requires $\braket{\tau_z}(x<0) = -\braket{\tau_z}(x>0)$, thus for perfect transmission $\braket{\tau_z}(x<0)= -\braket{M^\dagger_y\tau_z M_y}(x<0)$. 
%
Since $M_y =  e^{U_y/t} P_z(+) + e^{-U_y/t}P_z(-)$, in the limit $|U_y|>>|t|$ the spinors get projected on to eigenspinors of $\tau_z$ i.e. on to $\ket{\tau_z,+}$ ($\ket{\tau_z,-}$) for $U_y>0$ ($U_y<0$). This indicates spin-filtering along $z$. Due to SML, for $U_y>0$ ($U_y<0$) perfect transmission is only possible for $|k_z|>|k_0|$ ($|k_z|<|k_0|$), which agress with position of maxima in Fig \ref{fig:transm_prob}(c). 

For the $U_z\tau_z$ potential, the $M$-matrix satisfies $M^\dagger_z \tau_{x,z} M_z = \tau_{x,z}$. This prevents perfect transmission when $U_z\neq 0$, since the boundary matrix must change $\tau_z \rightarrow -\tau_z$ for perfect transmission. The only exception is when $\braket{\tau_z} = 0$, thus maxima is always located at $k_z = \pm k_0$ but magnitude is suppressed by $U_z$ as $|t_1|^2 = 1-|r_1|^2 = \text{sech}^2(U_z/t)$, see Fig. \ref{fig:transm_prob}(d). 



\paragraph{Spin-resolved Local Density of States.} 
Bound state contribution to LDOS at an energy, $E$, is obtained by integrating spin/particle density, $\braket{\psi|\tau_i|\psi}$, over the Fermi arcs, $\rho^{(b)}_i(\bm{r},E) = 1/(2\pi)^2 \int dk_y \int dk_z \delta(E - E_b) \braket{\tau_i}$. When the (Fermi) energy is at the Weyl nodes, bound state contribution to LDOS will be the only relevant contribution. It decays with a localization length $1/2\mu$, where $\mu = \sec (U_0/t) (k_0 \sin k_0 - E \sin (U_0/t))$ in the presence of only the electrostatic potential. 
For $U_0 = n\pi t$, the Fermi arcs are along $k_z$, and the spins of the bound states are oriented parallel or antiparallel to $y$-axis, as for $U_0 = 0$ in Figs. \ref{fig:bound_states}(a) and \ref{fig:bound_states}(c). For $U_0 \neq n\pi t$, both $\rho_y^{(b)}$ and $\rho_z^{(b)}$ can be non-zero with ratio $\rho^{(b)}_y/\rho^{(b)}_z = \cot(U_0/t) \text{sgn}(x)$. $\rho^{(b)}_y$ is even across the interface while $\rho^{(b)}_z$ is odd as expected from the chirality reversal, see SM \cite{SM}. 

Interface states, bound or scattering, are unaffected by magnetic potential $U_x\tau_x$. 
The potential $U_y\tau_y$ maintains finite spin-resolved LDOS and changes the shape of the fermi arcs, see SM \cite{SM}. The potential $U_z\tau_z$ maintains the shape and texture of the bound states for $U=0$, thus $\rho_y^{(b)}\neq 0$ while $\rho_z^{(b)}= 0$ vanishes. However, the potential also pushes the states to one side of the interface, see Fig. \ref{fig:ldos}(a). 

An interface, in addition to binding electrons, will cause Friedel oscillations in the LDOS from interference of incident and reflected electrons. 
The resultant contribution to LDOS is, $\rho_i^{(s)} = 1/(2\pi)^3 \sum_{proc} \sum_n\int d\bm{k} \delta(E-E_s) \braket{\psi|\tau_i|\psi}$, where $E_s = \epsilon \sqrt{k_x^2+k_y^2+(k_z-nk_0)^2\sin ^2k_0}$. The $U_z\tau_z$ potential leads to Friedel oscillations with a period $\pi t/E$ in both $\rho_y^{(s)}$ and $\rho_z^{(s)}$, and these carry symmetry signatures of the potential. The system has $\mathcal{R}_{z}(\pi) \mathcal{I}_{z\pm}$ symmetry,  where $\mathcal{R}_{i}(\pi)$ is 180$^\circ$ rotation about $i$-axis and $\mathcal{I}_{j\pm}$ is $j$-coordinate reversal about the node center, with $\pm$ corresponding to $n=\pm 1$. This makes $\rho_z^{(s)}$ even across the interface and $\rho_y^{(s)}$ odd across it, while $\rho_x^{(s)}$ vanishes, see  Fig. \ref{fig:ldos}(a) and SM for symmetry analysis. Similarly, in presence of only $U_y\tau_y$ potential, the system has $\mathcal{R}_{y}(\pi) \mathcal{I}_{z\pm}$ symmetry, thus $\rho_y^{(s)}$ is even while $\rho_z^{(s)}$ is odd across the interstitial layer, see SM \cite{SM}. 


\paragraph{Transport.} Longitudinal conductance through the interface in the ballistic limit is given by Landauer formula, $G = e^2/h \sum_{FS} |t_1|^2$, where sum is over all states on the Fermi surface. In absence of interstitial potentials, the chirality reversal reduces the conductance to $2/3$rd of the perfect transmission conductance $G_0 = \frac{k_F^2}{4\pi}\frac{e^2}{h}$. Electrostatic potential does not affect this value, see Fig. \ref{fig:ldos}(b) and SM \cite{SM}, because transmission maxima on the Fermi surface are only dislocated not suppressed by it. The potential $U_z\tau_z$ suppress the conductance as expected from the maxima value of $|t_1|^2\sim \text{sech}^2(U_z/t)$, and $U_x\tau_x$ leaves it unaffected. The potential $U_y\tau_y$ suppresses conductance overall, but leads to energy independent spin filtration along $z$ for moderate values, see  spin degree of polarization (SDOP) in Fig. \ref{fig:ldos}(c), where $SDOP_i = \int_{FS} |t_1|^2 \braket{\psi_t|\tau_j|\psi_t}/ \int_{FS} \braket{\psi_i|\psi_i}$ and $\psi_{t(i)}$  is the spinor part of the transmitted (incident) wavefunction. This high degree of spin-filtering occurs due to spin projection behavior of the magnetic potentials for the SML electrons.


In addition, there is a transverse ground state charge current due to $j_y \propto \mu t\braket{\tau_y}$ for WSM. This current was previously studied as axial current from linearly dispersing psuedo-Landau level \cite{Araki_2016,Grushin_2016} at a DW. Here we have focused on other aspects that were either unexplored or absent in DW systems.

\paragraph{Discussion and Conclusion.} We showed using a low-energy model that interstitial  layer potentials can tune the states near a CRPI in WSMs. The electrostatic and in-plane magnetic potential perpendicular to chiral node split direction determine the shape of Fermi-arc-like bound states. The other in-plane component, $U_z\tau_z$, determines asymmetry of local density of states across the interface. 
The effect of $U_0$ is in-plane rotation, while effect of $U_y\tau_y$ and $U_z\tau_z$ can be understood as spin-projections along $z$ and $y$-axes respectively. Transmission probabilities of bulk states confirm this behavior. We showed that $U_y \tau_y$ can control spin-filtering of electrons across the interface, and combined with $U_z\tau_z$ and $U_0\tau_0$ it can control their overall transmission. 
Inability of $U_x\tau_x$ to affect electron transport allows one to switch on or off the effect of magnetic potentials by changing their ratio, similar to line interfaces in topological insulators \cite{Thareja_2023}. This opens new possibilities for applied physics.

{Recently, phase diagrams for photonic counterparts of CRPIs have been explored for type-I \cite{Song_2023} and type-II \cite{Song_2024} Weyl heterostructures. A generalization of our work to these photonic systems would be of interest to applied physics. Our work also motivates the exploration of the effects of more complex magnetic textures, such as skyrmions \cite{Wu_2020}, on WSMs electronic states. } 



%

\begin{acknowledgements}
This material is based upon work supported by the Air Force Office of Scientific Research under award number FA9550-23-1-0132.
J.G. and G.P. acknowledges support from the Max Planck Society through the Max Planck Partner Group Programme.
\end{acknowledgements}


%

\end{document}